\documentclass[showpacs,a4paper,superscriptaddress]{revtex4}

\usepackage{graphicx}
\usepackage{graphics}
\usepackage{xspace}
\usepackage{amssymb}
\usepackage{amsmath}
\usepackage{latexsym}
\usepackage{mathrsfs}
\usepackage{pstricks}

\newcommand{\ii}{{\rm i}}
\newcommand{\de}{{\rm\,d}}
\newcommand{\e}{{\rm e}}

\newcommand{\im}{\,{\rm Im}\,}
\newcommand{\tr}{\mbox{Tr}\,}
                 
\newcommand{\st}{{\scriptscriptstyle T}}
\newcommand{\sa}{{\scriptscriptstyle A}}
\newcommand{\sss}{{\scriptscriptstyle S}}

\newcommand{\g}{\gamma}
\newcommand{\sig}{\sigma}
\newcommand{\eps}{\epsilon}

\newcommand{\pslash}{\rlap{/} p}
\newcommand{\qslash}{\rlap{/} q}
\newcommand{\kslash}{\rlap{/} k}
\newcommand{\lslash}{\rlap{/} l}

\begin{document}
 

\title{
Collins fragmentation function from gluon rescattering}

\author{Alessandro Bacchetta}
\email{alessandro.bacchetta@physik.uni-regensburg.de}
\affiliation{Institut f\"ur Theoretische Physik, Universit\"at Regensburg,
D-93040 Regensburg, Germany}

\author{Andreas Metz}
\email{andreas.metz@tp2.ruhr-uni-bochum.de}
\affiliation{Institut f\"ur Theoretische Physik II, Ruhr-Universit\"at Bochum,
D-44780 Bochum, Germany}

\author{Jian-Jun Yang}
\email{jianjun.yang@physik.uni-regensburg.de}
\affiliation{Institut f\"ur Theoretische Physik, Universit\"at Regensburg,
D-93040 Regensburg, Germany}

\begin{abstract}
We estimate the Collins fragmentation function by
 introducing the effect of gluon
rescattering in a model calculation of the fragmentation process. 
We include all necessary diagrams to the one-loop level and
compute the nontrivial phases giving rise to the Collins function. We
compare our results to the ones obtained from pion rescattering.
We conclude that three out of four
one-loop diagrams give sizeable  
contributions to the Collins function, and
that the effect of gluon rescattering 
has a magnitude comparable to that of pion rescattering, but has opposite sign.
\end{abstract}


\maketitle

\section{Introduction}

The Collins fragmentation function~\cite{Collins:1993kk} 
is an example of
how the orientation of the quark spin can influence the direction of
emission of hadrons in the fragmentation process. 
The existence and the features of this function are related to important
questions of nonperturbative QCD, such as 
the role of chiral symmetry and color gauge
invariance in the hadronization process.
The Collins function is
believed to be at the origin of single-spin asymmetries in hard hadronic
reactions~\cite{Adams:1991cs,Bravar:1996ki,Bravar:2000ti,Airapetian:2000tv,Airapetian:2001eg,Avakian:2003pk},
 which lead to attempts of estimating it 
from phenomenology~\cite{Artru:1997bh,Anselmino:1998yz,Anselmino:1999pw,Boglione:2000jk,Efremov:2001cz}.

The Collins function is a so-called  T-odd entity.
T-odd functions typically
require the interference 
between two amplitudes with different imaginary parts to exist.
Perturbative calculations of fragmentation functions in quantum field theories 
can provide -- through loop corrections -- the necessary nontrivial phases in
the fragmentation amplitude~\cite{Bacchetta:2001di}.

The Collins fragmentation function has been estimated in a chiral invariant
approach where the effective degrees of freedom are 
constituent quarks and pions, coupled via a pseudovector 
interaction~\cite{Bacchetta:2002tk}. In order to generate the 
required phases, one-loop corrections to the quark propagator and vertex have
been included.

In the meanwhile, gluon loop corrections in distribution functions 
have been investigated. 
In the context of a spectator model of the nucleon, it has been shown that
the exchange of a gluon between the struck quark and the target spectators
gives rise to T-odd distribution functions~\cite{Brodsky:2002cx,Collins:2002kn}. 
This has
been interpreted as the one-gluon approximation to the gauge link, which is
 included 
in the definition of parton distribution
functions and insures their color gauge 
invariance~\cite{Collins:2002kn,Ji:2002aa,Belitsky:2002sm,Boer:2003cm}. 
It was soon realized that gauge link effects have different signs in 
semi-inclusive DIS and Drell-Yan scattering~\cite{Collins:2002kn,Brodsky:2002rv}, implying
different signs for T-odd distribution functions in the two processes. This is 
for the first time an example of violation of universality of the distribution 
functions, though it consists simply of a sign change.
Gluon rescattering has been  used thereafter to estimate T-odd distribution
functions~\cite{Boer:2002ju,Gamberg:2003ey}.  

As in the distribution functions, the gauge link can generate nontrivial
phases and T-odd effects in the fragmentation functions as well. This was
analyzed in Ref.~\cite{Metz:2002iz}, where it was shown in
particular that one-gluon contributions to the gauge link 
don't
change sign in semi-inclusive DIS and in $e^+e^-$ annihilation.

In this work, we model the tree-level fragmentation process in the same way as 
in Ref.~\cite{Bacchetta:2002tk}. To generate T-odd effects, instead of
introducing pion-loop corrections, we include gluon rescattering corrections. 
We take into consideration all possible one-loop amplitudes,
including those that arise from the gauge link.
We compute the Collins function and its first two moments, which can be
experimentally accessed in single-spin asymmetries in semi-inclusive DIS, as
well as in $e^+e^-$ annihilation.
A similar calculation has been presented very recently by Gamberg et
al.~\cite{Gamberg:2003eg}, using a pseudoscalar coupling complemented 
with a Gaussian form factor to model the tree
level fragmentation. However, in that work
only one of the four possible gluon rescattering
contributions has been taken into account.

\section{Calculation of the Collins function}

We use the following definition of the Collins function
$H_1^\perp$~\cite{Levelt:1994np,Mulders:1996dh}:
\begin{align} 
\frac{\eps_{\st}^{ij} k_{\st j}}{m_{\pi}}
	      \, H_1^{\perp}(z,z^2 k^2_{\st}) 
&=            \tr[ \Delta(z,k_\st)\, \ii \sig^{i-}\g_5] \,.  
\label{e:col1}
\end{align} 
The correlation function $\Delta(z,k_\st)$ can be written as~\cite{Boer:2003cm}
\begin{equation} \begin{split} 
\Delta(z,k_\st) &= \frac{1}{4z} \int \de k^+ \;\Delta
(k,p)\,\bigg|_{k^-=p^-/z} \\
	 & =\sum_X \, \int
        \frac{\de\xi^+ \de^2\xi_\st}{4z (2\pi)^{3}}\; \e^{+\ii k \cdot \xi}
       \langle 0|
{\cal L}_{[-\infty^+, \xi^+]}
\,\psi(\xi)|\pi, X\rangle 
\langle \pi, X|
             \bar{\psi}(0)\,{\cal L}_{[0^+, -\infty^+]}
|0\rangle \bigg|_{\xi^-=0}\,,    
\label{e:delta}
\end{split} \end{equation}  
where the notation ${\cal L}_{[a,b]}$ indicates a straight gauge link running 
from $a$ to $b$. In this work, we will use the Feynman gauge.  
In the case of transverse-momentum independent fragmentation
functions, i.e., after integrating over $k_\st$, 
by choosing a light-cone gauge the link can be
reduced to unity. However, in the case of transverse-momentum dependent
functions -- as the Collins function -- the 
gauge link cannot be
neglected and becomes one of the possible sources of nontrivial phases in the 
fragmentation amplitude and thus of T-odd fragmentation
functions~\cite{Metz:2002iz}. Note that in case of transverse-momentum
dependent 
functions 
with lightlike Wilson lines, divergencies can
appear~\cite{Collins:1982uw,Collins:2003fm}. However, for our study this
problem does not show up.

To model the tree-level fragmentation process, we make use of the chiral
invariant effective model of Manohar and Georgi~\cite{Manohar:1984md}.
The unpolarized fragmentation function in this model
reads~\cite{Bacchetta:2002tk} 
\begin{equation} \begin{split} 
 D_1(z,z^2 k^2_\st) & = 
 \frac{1}{z} \frac{g_\sa^2}{4 F_\pi^2}
 \frac{1}{16\pi^3} 
\bigg( 1 - 4\frac{1-z}{z^2}
 \frac{m^2 m_\pi^2}{\left[k_\st^2 +m^2 + m_\pi^2(1-z)/z^2\right]^2} \bigg). 
\label{e:d1tree}
\end{split} \end{equation}

To obtain a nonzero Collins function, we have to compute one-loop
contributions. In contrast to what was done in our previous
calculation~\cite{Bacchetta:2002tk}, 
instead of introducing pion loops, we now consider the effect of including
gluon rescattering.

	\begin{figure}
	\centering
	\begin{tabular}{ccccc}
        \includegraphics[width=5 cm]{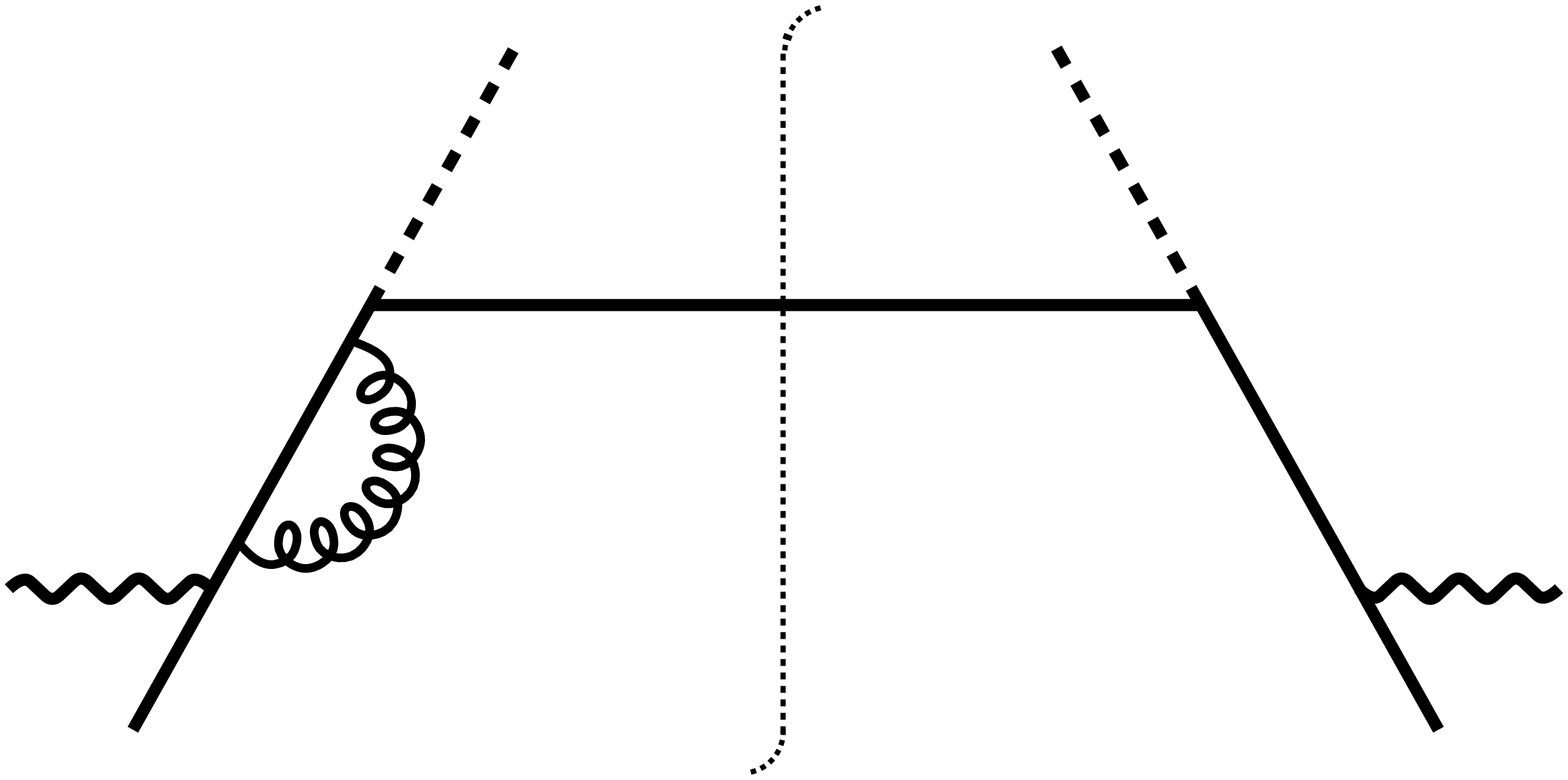}& \makebox[0.5cm]{}&
        \includegraphics[width=5 cm]{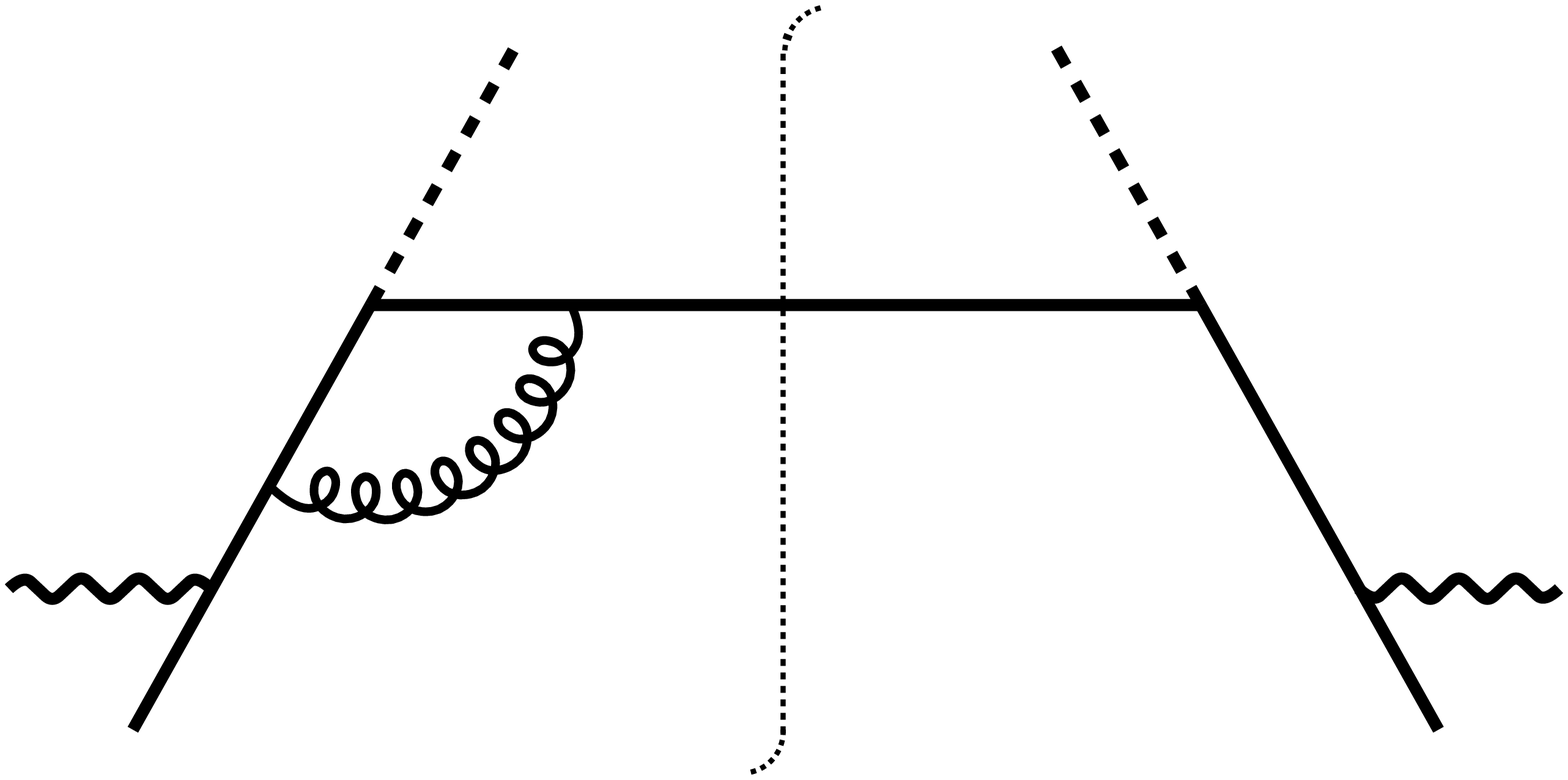} \\ \\
		(a) & &(b)\\ \\
        \parbox[m]{5cm}{\includegraphics[width=5 cm]{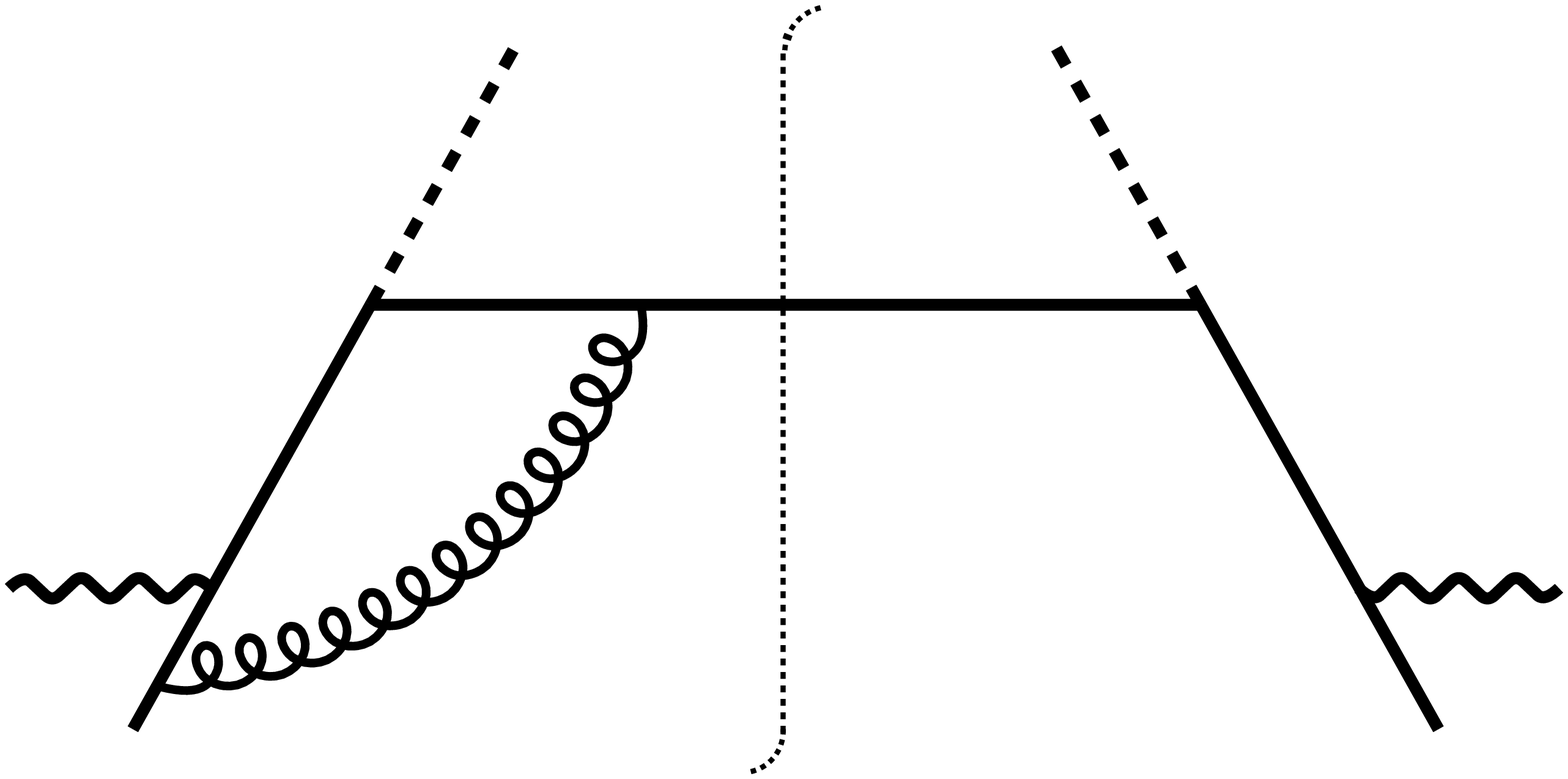}} &&
        \parbox[m]{5cm}{\includegraphics[width=5 cm]{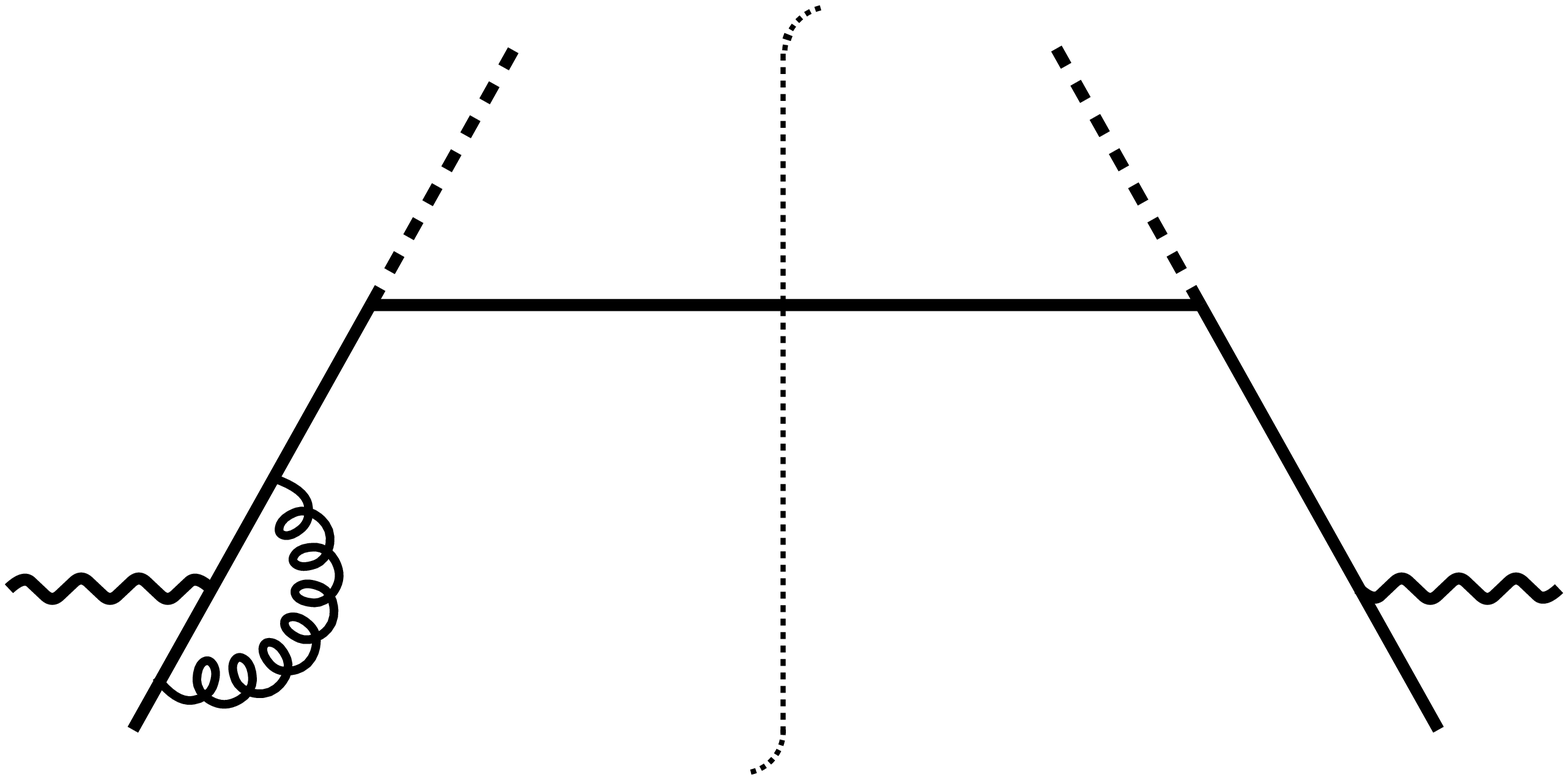}}
	 &\makebox[0.5cm]{}&+ H.~c.\\ \\
	(c) & &(d)\\ \\
	\end{tabular}
        \caption{Single gluon-loop corrections to the fragmentation of a quark
		 into a pion contributing to the Collins function.
                 The Hermitian conjugate diagrams (H.~c.) are not shown
                 explicitly.}
        \label{f:1loop}
        \end{figure}

In Fig.~\ref{f:1loop} we illustrate the Feynman diagrams involved in the
calculation of the Collins function. Diagrams (a) and (b) -- 
the self-energy and pion vertex corrections -- 
have analogous contributions in the pion-loop case. 
However, due to gauge invariance we have to take into consideration also
diagrams (c) and (d), which we will henceforth call the box and
photon vertex corrections. Despite their appearance, these contributions do not 
break factorization, since they can be interpreted as contributions to the 
gauge link, which is included 
in
Eq.~(\ref{e:delta})~\cite{Collins:2002kn,Ji:2002aa,Belitsky:2002sm,Boer:2003cm}.
In fact, diagram (d) contributes not only to the gauge link of the
fragmentation function. However, since in this specific calculation its
contribution turns out to vanish, we will avoid considering this issue.

\begin{figure}
	\centering
	\begin{tabular}{ccccccc}
        \includegraphics[width=3.5 cm]{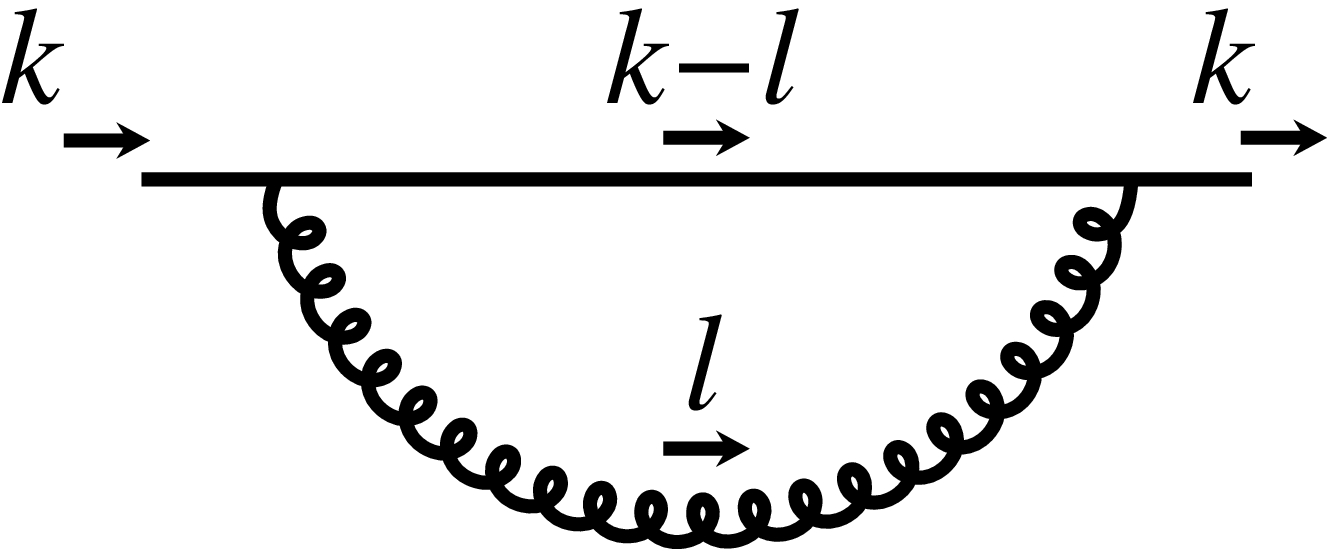}&\makebox[0.5cm]{} &
        \includegraphics[width=3.5 cm]{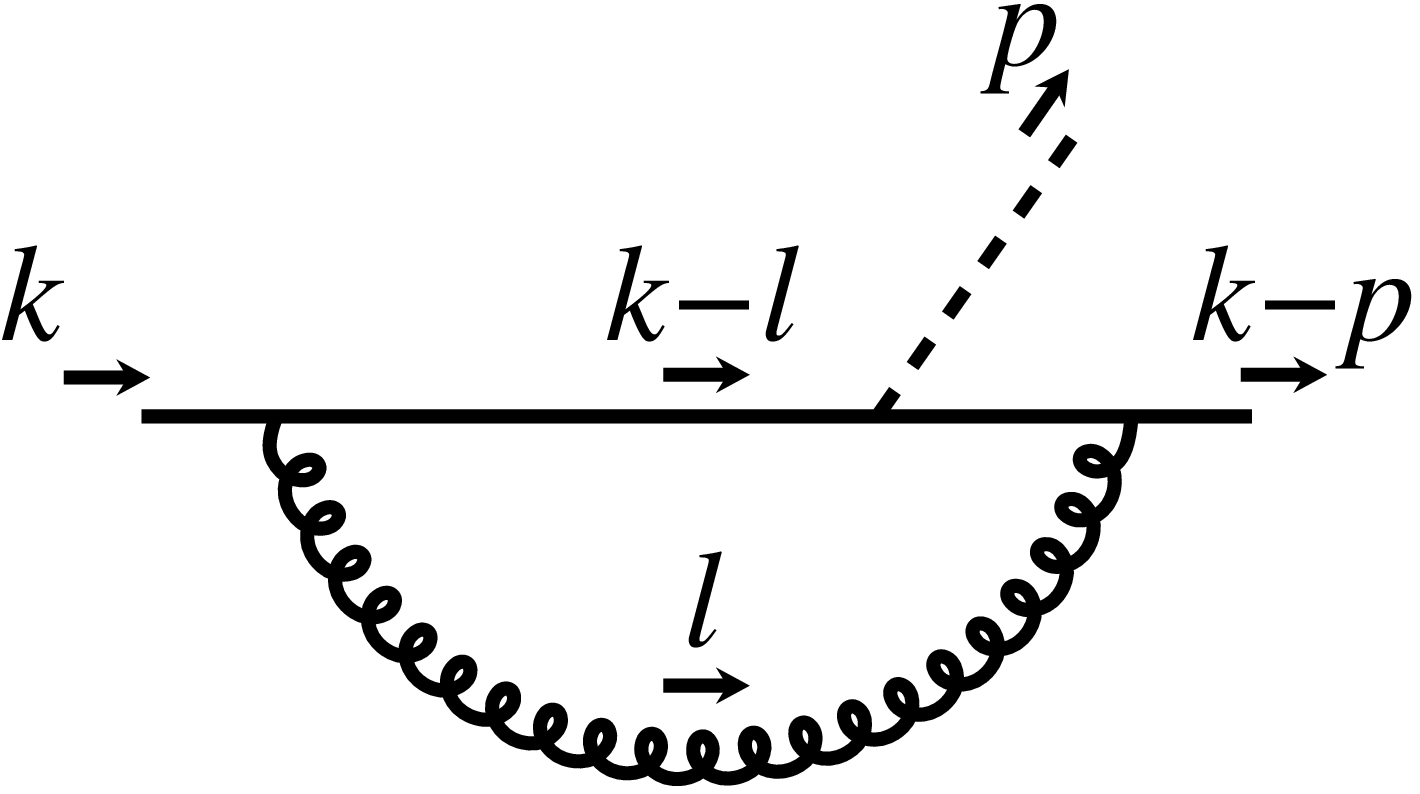}&\makebox[0.5cm]{} &
        \includegraphics[width=3.5 cm]{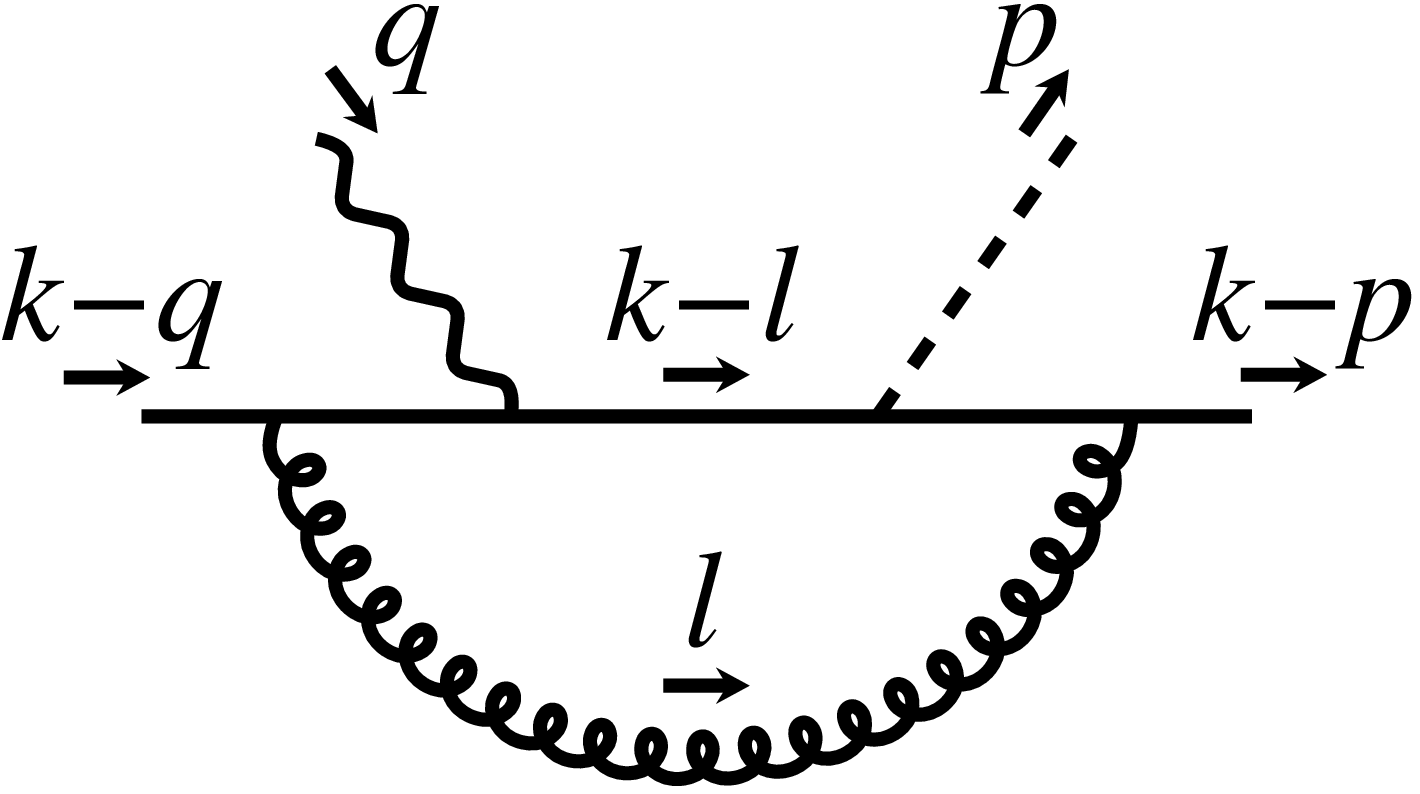}&\makebox[0.5cm]{} &
        \includegraphics[width=3.5 cm]{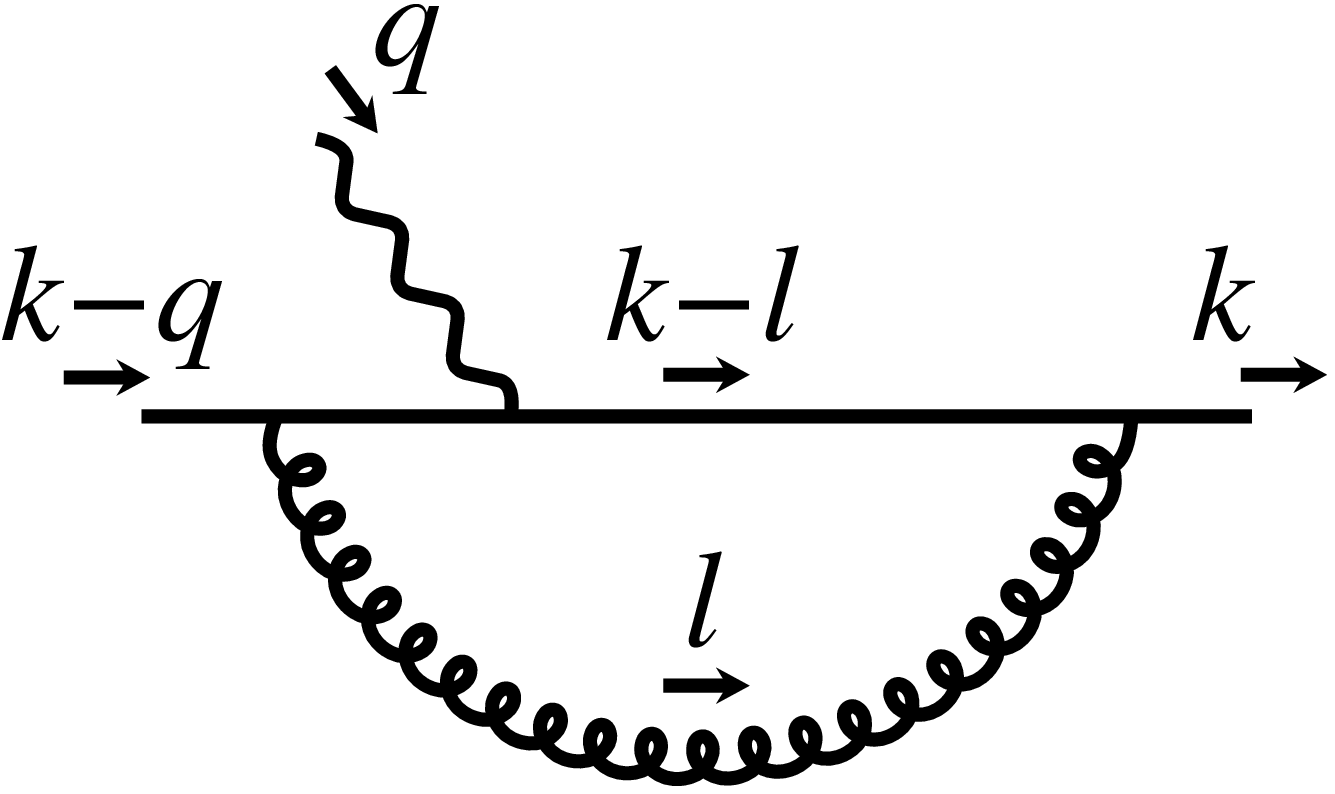} \\ \\
	 $- \ii\, \Sigma(k)$ & & $\Gamma(k,p)$ & & $\ii\, \Xi^{\mu}(k,q,p)$ & &
				$\Phi^{\mu}(k,q)$       
	\end{tabular}  
	\caption{One-loop self-energy, pion vertex, box, and photon vertex 
	 corrections.}
        \label{f:sigmagamma}
        \end{figure}

In all these diagrams, the gluon loops can  give rise to nontrivial phases
to the scattering amplitude. Through the interference with the
tree-level amplitude, these contributions generate nonzero T-odd fragmentation 
functions, such as the Collins function.
The portions of the diagrams relevant for the generation of imaginary parts are
sketched in Fig.~\ref{f:sigmagamma}. They can be expressed analytically as
\begin{align}
 \Sigma (k) &= - \ii g_\sss^2\,C_F\,
 \int \frac{\de^4 l}{(2 \pi)^4} \; 
 \frac{\gamma_{\rho}\, (\kslash - \lslash + m) \, \gamma^{\rho}}
 {[(k-l)^2 - m^2]\,[l^2 -m_{g}^2]} \,,
\\
\begin{split}
\Gamma (k,p) &= - \ii \frac{g_\sa}{2 F_\pi}  g_\sss^2\,C_F\,
 \int \frac{\de^4 l}{(2\pi)^4}\;
\frac{\gamma_{\rho} \, (\kslash - \pslash - \lslash + m)}
 {[(k - p - l)^2 - m^2]} \,\frac{\pslash \,\gamma_5 (\kslash - \lslash + m) \, \gamma^{\rho}}
 {[(k - l)^2 - m^2][l^2 - m_{g}^2]} \,,
\end{split}\\
\begin{split}
\Xi^{\mu}(k,q,p) &=- \ii \frac{g_\sa}{2 F_\pi}  g_\sss^2\,C_F\,
 \int \frac{\de^4 l}{(2\pi)^4}\;
\frac{\gamma_{\rho} \, (\kslash - \pslash - \lslash + m)\,\pslash \,\gamma_5}
 {[(k - p - l)^2 - m^2][l^2 - m_{g}^2]} \,\frac{(\kslash - \lslash + m) \, 
\gamma^{\mu}\,(\kslash - \qslash - \lslash + m)\,\gamma^{\rho}}
 {[(k - l)^2 - m^2][(k - q - l)^2 - m^2]} \,,
\end{split}\\
\begin{split}
\Phi^{\mu} (k,q) &= - \ii  g_\sss^2\,C_F\,
 \int \frac{\de^4 l}{(2\pi)^4}\;
\frac{\gamma_{\rho} \, (\kslash - \lslash + m)\,\gamma^{\mu}}
 {[(k - q - l)^2 - m^2]} \,\frac{
(\kslash - \qslash - \lslash + m)\, \gamma^{\rho}}
 {[(k - l)^2 - m^2][l^2 - m_{g}^2]} \,,
\end{split}
\end{align} 
where $C_F=4/3$.
We choose the kinematics in the following way (in light-cone coordinates):
\begin{align}
q& =\left[\frac{Q}{\sqrt{2}}, - \frac{Q}{\sqrt{2}}, 0_\st\right], & k&=\left[\frac{Q}{\sqrt{2}},
 \frac{k^2 + k_T^2}{Q \sqrt{2}}, k_\st\right],
& p &= \left[z \frac{Q}{\sqrt{2}},
 \frac{m_{\pi}^2}{z Q \sqrt{2}}, 0_\st\right].
\end{align} 

To identify the contributions to the Collins function, 
we write down the explicit expressions for the cut diagrams
of Fig.~\ref{f:1loop} and we apply the definition of
the Collins function given in Eq.~(\ref{e:col1}).

It turns out that only some specific elements of the 
imaginary parts of the diagrams in Fig.~\ref{f:sigmagamma}
contribute to the Collins function. 
For simplicity we will denote
them as $\im \sigma$, $\im \gamma$, $\im \xi$, $\im \phi$. The formula of the
Collins function can be written then as 
\begin{equation} \begin{split} 
H_1^{\perp}(z,z^2 k_\st^2) & = 
 \frac{g_\sa^2 \, m_{\pi}}{16 \pi^3 F_{\pi}^2} \,m \, k^- \int d k^+
	\delta((k-p)^2 - m^2) \frac{1}{k^2 - m^2}
\Bigl(\im \sigma + \im \gamma + \im \xi + \im \phi \Bigr)
\\
&=  \frac{g_\sa^2}{32 \pi^3 F_\pi^2} \frac{m_\pi}{1-z} \frac{m}{k^2 - m^2} 
\Bigl(\im \sigma + \im \gamma + \im \xi + \im \phi \Bigr)
\bigg|_{k^2 = k_{T}^2 \frac{z}{1-z} + \frac{m^2}{1-z} + \frac{m_{\pi}^2}{z}} \,.
\end{split} \end{equation}   

Considering only the leading power in $Q$ and assuming the gluons to be
massless, the calculation of the four diagrams yields
\begin{align} 
\im \sigma & =  -
 \frac{\alpha_s}{2\pi} \, C_F \,
 \biggl(3 - \frac{m^2}{k^2} \biggr)  I_{1,g},
\\ 
\im \gamma & =
 \frac{\alpha_s}{2\pi} \, C_F \,
 \biggl[ \bigg(1 + \frac{m^2}{k^2} \biggr)  I_{1,g} 
      + 4 m_\pi^2 \, I_{2,g} \bigg],
\\ 
\begin{split}
\im \xi & =
- \frac{\alpha_s}{\pi} \, C_F \, 
 \bigg \{ 2\, I_{1,g} + 2 Q^2\,\Bigl( (k^2 - m^2)(1-z) I_{4,g} + I_{3,g}\Bigr)
+2 \Bigl(2 z m^2 - (1-z)(k^2-m^2)\Bigr)
\\ &\qquad
\times\frac{1}{2 z^2 k_\st^2}
\left[z\,Q^2\,\Bigl((k^2 - m^2)(1-z) I_{4,g} + I_{3,g}\Bigr) - \Bigl(z(k^2 - m^2 +m_{\pi}^2) -2m_{\pi}^2 \Bigr) I_{2,g}\right]
      \bigg\}, 
\end{split}\\
\im \phi &= 0, 
\end{align} 
where the integrals introduced above correspond to
\begin{align} 
\begin{split} 
I_{1,g} & = 
 \int d^4 l \; \delta(l^2) \, \delta((k-l)^2 - m^2)
\;
 = \frac{\pi}{2 k^2}\;(k^2 -m^2)\;\theta(k^2 - m^2), 
\end{split}\\
\begin{split}
I_{2,g} & =
 \int d^4 l \; \frac{\delta(l^2) \, \delta((k-l)^2 - m^2)}
 {(k - p - l)^2 - m^2}
\;
= 
- \frac{\pi}
 {2\sqrt{\lambda(k^2,m^2,m_{\pi}^2)}}
\\
& \qquad \times 
 \ln \biggl( 1 + \frac{2 \sqrt{\lambda(k^2,m^2,m_{\pi}^2)}}
  {k^2- m_{\pi}^2+ m^2- \sqrt{\lambda(k^2,m^2,m_{\pi}^2)}} \biggr)
 \; \theta(k^2 - m ^2),
\end{split} \\
\begin{split}
I_{3,g} & = \int d^4 l \; \frac{\delta(l^2) \, \delta((k-l)^2 - m^2)}
 {(k - q - l)^2 - m^2}
\;
= 
- \frac{\pi}{Q^2} \ln{\frac{Q^2}{m \sqrt{k^2}}}\;\theta(k^2 - m^2),
\end{split} \\
\begin{split}	
I_{4,g} & = \int d^4 l \; \frac{\delta(l^2) \, \delta((k-l)^2 - m^2)}
 {\left[(k - p - l)^2 - m^2 \right] \left[(k - q - l)^2 - m^2\right]}
\\
& =
\frac{\pi}{Q^2} 
\frac{1}{(k^2 - m^2)(1-z)} 
\biggl(\ln \frac{Q^2}{m \sqrt{k^2}}
 + \ln \frac{ \sqrt{k^2} (1-z)}{m} \biggr)\;\theta(k^2 - m^2).
\end{split} 
\end{align}  
In the previous formulae, we made use of the K\"allen function, 
$\lambda(k^2,m^2,m_{\pi}^2)=[k^2 -(m+m_{\pi})^2][k^2 -(m-m_{\pi})^2]$.
Note that the final result for the Collins function is independent of $Q^2$.

\section{Numerical results}

We perform the numerical
integration over the transverse momentum according to
 \begin{equation} 
D_1(z)= \pi \int_0^{K^{2}_{\st\,{\rm max}}} \de K^2_{\st}\; D_1(z,K^2_{\st}),
\end{equation} 
where ${K}_{\st}=- z {k}_{\st}$.
We impose a cutoff on the virtuality of the fragmenting quark, so that 
$k^2 \le \mu^2$. Due to the kinematics, this choice fixes
the upper limit of the $K^2_{\st}$ integration to 
\begin{equation} 
K^{2}_{\st\,{\rm max}}=z \, (1-z)\,\mu^2 -z\,m^2-(1-z)\,m_\pi^2 \,.
\end{equation} 
For the numerical computations, we use the following values of the parameters
of our model:
\begin{equation} 
m_q= 0.3 \; {\rm GeV}, \qquad g_{\sa}=1, \qquad \mu^2=1\; {\rm GeV}^2, 
\qquad \alpha_s=0.3 \,.
 \label{e:param}
\end{equation} 
The choice of the first three parameters has been extensively discussed in
Ref.~\cite{Bacchetta:2002tk}.  The choice of the strong coupling 
constant $\alpha_s$ corresponds to a reasonable value for $Q^2\approx 1 \;{\rm
GeV}^2$. 

        \begin{figure}[b]
        \centering
        \rput(-0.4,3.1){\rotatebox{90}
	{{\boldmath $H_1^{\perp (1/2)}/D_1$ }} }
	\rput(4.5,-0.3){\boldmath $z$}
	\includegraphics[width=8cm]{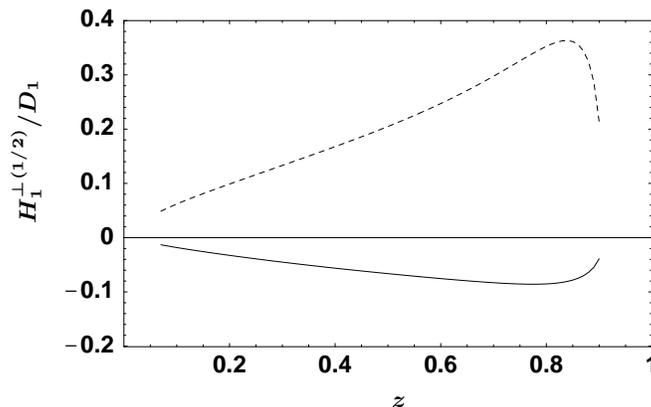}
        \caption{Result for $H_1^{\perp (1/2)}/D_1$ 
	including only the
        self-energy and vertex  
	gluon-loop corrections
	(solid line) and comparison with 
	the result of the pion-loop model
	(dashed line) from Ref.~\cite{Bacchetta:2002tk}.}
	\label{f:ratiomom0a}
        \end{figure}

\begin{figure}
        \centering
        \rput(-0.4,3.1){\rotatebox{90}
	{{\boldmath $H_1^{\perp (1/2)}/D_1$ }} }
	\rput(4.5,-0.3){\boldmath $z$}
	\includegraphics[width=8cm]{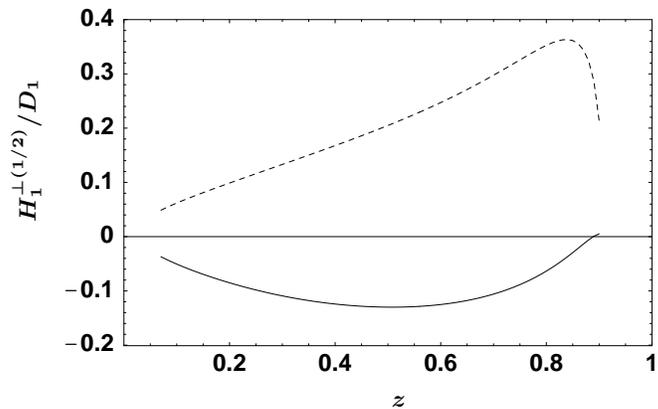}
        \caption{Result for $H_1^{\perp (1/2)}/D_1$ 
	including all gluon-loop contributions
	(solid line) and comparison with 
	the result of the pion-loop model
	(dashed line) from Ref.~\cite{Bacchetta:2002tk}.}
	\label{f:ratiomom0}
        \end{figure}

In Fig.~\ref{f:ratiomom0a} and \ref{f:ratiomom0} we plot the ratio
\begin{equation}	
 \frac{H_1^{\perp
(1/2)}(z)}{D_1(z)} \equiv
\frac{\pi}{D_1(z)} \int \de K_{\st}^2\, \frac{|{K}_{\st}|}{2 z m_{\pi}}\,
H_1^{\perp}(z,K_{\st}^2) \,,
\label{e:ratiomom0}
\end{equation} 
which enters the unweighted transverse single spin
asymmetries for pion production in
semi-inclusive DIS~\cite{Mulders:1996dh,Bacchetta:2002tk}.  
In
Fig.~\ref{f:ratiomom0a} we show only the contributions of the self-energy and
pion vertex corrections, diagrams (a) and (b) of Fig.~\ref{f:1loop}. They have 
a direct correspondence to the pion-loop case. As shown by the plot, these
contributions are smaller and have an opposite sign compared to the pion-loop ones. 
In
Fig.~\ref{f:ratiomom0a} we show the sum of all gluon-loop 
contributions, i.e.\ all diagrams of Fig.~\ref{f:1loop}, including in
particular the box diagram, which contributes to the gauge link and 
has no analogous term in the pion case. 

        \begin{figure}
        \centering
        \rput(-0.4,3.1){\rotatebox{90}
	{{\boldmath $H_1^{\perp (1)}/D_1$ }} }
	\rput(4.5,-0.3){\boldmath $z$}
        \includegraphics[width=8cm]{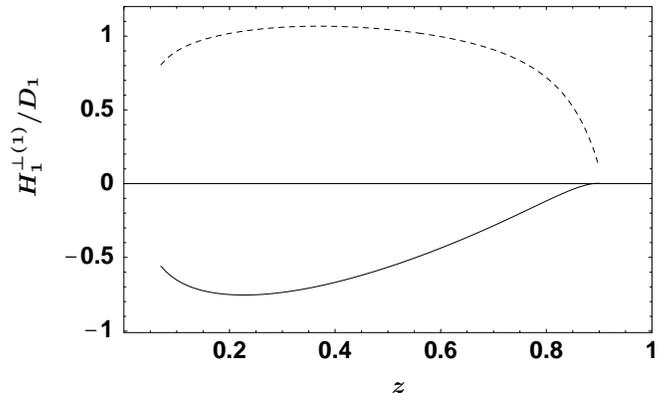}
        \caption{Result for $H_1^{\perp (1)}/D_1$ including all
	gluon-loop contributions
	(solid line) and comparison with 
	the result of the pion-loop model
	(dashed line) from Ref.~\cite{Bacchetta:2002tk}. } 
	\label{f:ratiomom}
        \end{figure}

In Fig.~\ref{f:ratiomom} we plot the ratio
\begin{equation}	
 \frac{H_1^{\perp(1)}(z)}{D_1(z)} \equiv
\frac{\pi}{D_1(z)} \int \de K_{\st}^2\, \frac{K_{\st}^2}{2 z^2 m_{\pi}^2}\,
H_1^{\perp}(z,K_{\st}^2) \,,
\end{equation} 
which typically enters weighted transverse spin
asymmetries in
semi-inclusive DIS~\cite{Kotzinian:1997wt,Boer:1998nt,Bacchetta:2002tk}.

\section{Conclusions}

In this work we have calculated the Collins function for pions describing the
fragmentation process at tree level by means of a chiral invariant effective
theory, and generating the required nontrivial phases 
by means of gluon rescattering.
We have computed all necessary diagrams contributing to the Collins function
at the one-loop level, including the ones involved in the gauge link. 
Out of the four diagrams we considered, one gives no contribution, 
while the other three have similar magnitudes and
cannot be neglected. A word of caution is in order at this point: we have
performed a perturbative expansion in $\alpha_S$ in a regime where its
justification can be questioned.
Moreover, we emphasize once again that this is only a model
  calculation, whose reliability might be very limited. 
Nevertheless, we believe that it could give 
an indication of the influence of the gluon dynamics on the Collins
function. 

We compared our results with those formerly obtained in 
Ref.~\cite{Bacchetta:2002tk}, where pion rescattering was considered as a 
possible source of nontrivial phases. The two approaches are identical at
tree level, allowing a clear comparison of the differences at the one-loop 
level. 
We have found that the gluon rescattering mechanism produces a Collins
function with opposite sign compared to the pion rescattering mechanism. The
two effects are similar in magnitude, except perhaps at high $z$, where
pion rescattering dominates.
Cancellations between these two competing mechanism could decrease the
experimental asymmetries.


\begin{acknowledgments}
The work of A.~B.\ has been
supported by the TMR network HPRN-CT-2000-00130, the work of A.~M.\ by
the Sofia Kovalevskaya
Programme of the Alexander von Humboldt Foundation, and the work of J.~Y.\ by
the Alexander von Humboldt Foundation and by the
Foundation for University Key Teacher of the
Ministry of Education (China).
 
\end{acknowledgments}


\bibliographystyle{apsrev}
\bibliography{mybiblio}

\begin{thebibliography}{31}
\expandafter\ifx\csname natexlab\endcsname\relax\def\natexlab#1{#1}\fi
\expandafter\ifx\csname bibnamefont\endcsname\relax
  \def\bibnamefont#1{#1}\fi
\expandafter\ifx\csname bibfnamefont\endcsname\relax
  \def\bibfnamefont#1{#1}\fi
\expandafter\ifx\csname citenamefont\endcsname\relax
  \def\citenamefont#1{#1}\fi
\expandafter\ifx\csname url\endcsname\relax
  \def\url#1{\texttt{#1}}\fi
\expandafter\ifx\csname urlprefix\endcsname\relax\def\urlprefix{URL }\fi
\providecommand{\bibinfo}[2]{#2}
\providecommand{\eprint}[2][]{\url{#2}}

\bibitem[{\citenamefont{Collins}(1993)}]{Collins:1993kk}
\bibinfo{author}{\bibfnamefont{J.~C.} \bibnamefont{Collins}},
  \bibinfo{journal}{Nucl. Phys.} \textbf{\bibinfo{volume}{B396}},
  \bibinfo{pages}{161} (\bibinfo{year}{1993}),
  \eprint[http://arXiv.org/abs]{hep-ph/9208213}.

\bibitem[{\citenamefont{Adams et~al.}(1991)}]{Adams:1991cs}
\bibinfo{author}{\bibfnamefont{D.~L.} \bibnamefont{Adams}} \bibnamefont{et~al.}
  (\bibinfo{collaboration}{FNAL-E704}), \bibinfo{journal}{Phys. Lett.}
  \textbf{\bibinfo{volume}{B264}}, \bibinfo{pages}{462} (\bibinfo{year}{1991}).

\bibitem[{\citenamefont{Bravar et~al.}(1996)}]{Bravar:1996ki}
\bibinfo{author}{\bibfnamefont{A.}~\bibnamefont{Bravar}} \bibnamefont{et~al.}
  (\bibinfo{collaboration}{Fermilab E704}), \bibinfo{journal}{Phys. Rev. Lett.}
  \textbf{\bibinfo{volume}{77}}, \bibinfo{pages}{2626} (\bibinfo{year}{1996}).

\bibitem[{\citenamefont{Bravar}(2000)}]{Bravar:2000ti}
\bibinfo{author}{\bibfnamefont{A.}~\bibnamefont{Bravar}}
  (\bibinfo{collaboration}{Spin Muon}), \bibinfo{journal}{Nucl. Phys.}
  \textbf{\bibinfo{volume}{A666}}, \bibinfo{pages}{314} (\bibinfo{year}{2000}).

\bibitem[{\citenamefont{Airapetian et~al.}(2000)}]{Airapetian:2000tv}
\bibinfo{author}{\bibfnamefont{A.}~\bibnamefont{Airapetian}}
  \bibnamefont{et~al.} (\bibinfo{collaboration}{HERMES}),
  \bibinfo{journal}{Phys. Rev. Lett.} \textbf{\bibinfo{volume}{84}},
  \bibinfo{pages}{4047} (\bibinfo{year}{2000}),
  \eprint[http://arXiv.org/abs]{hep-ex/9910062}.

\bibitem[{\citenamefont{Airapetian et~al.}(2001)}]{Airapetian:2001eg}
\bibinfo{author}{\bibfnamefont{A.}~\bibnamefont{Airapetian}}
  \bibnamefont{et~al.} (\bibinfo{collaboration}{HERMES}),
  \bibinfo{journal}{Phys. Rev.} \textbf{\bibinfo{volume}{D64}},
  \bibinfo{pages}{097101} (\bibinfo{year}{2001}),
  \eprint[http://arXiv.org/abs]{hep-ex/0104005}.

\bibitem[{\citenamefont{Avakian}(2003)}]{Avakian:2003pk}
\bibinfo{author}{\bibfnamefont{H.}~\bibnamefont{Avakian}}
  (\bibinfo{collaboration}{CLAS}) (\bibinfo{year}{2003}),
  \eprint{hep-ex/0301005}.

\bibitem[{\citenamefont{Artru et~al.}(1997)\citenamefont{Artru, Czyzewski, and
  Yabuki}}]{Artru:1997bh}
\bibinfo{author}{\bibfnamefont{X.}~\bibnamefont{Artru}},
  \bibinfo{author}{\bibfnamefont{J.}~\bibnamefont{Czyzewski}},
  \bibnamefont{and} \bibinfo{author}{\bibfnamefont{H.}~\bibnamefont{Yabuki}},
  \bibinfo{journal}{Z. Phys.} \textbf{\bibinfo{volume}{C73}},
  \bibinfo{pages}{527} (\bibinfo{year}{1997}),
  \eprint[http://arXiv.org/abs]{hep-ph/9508239}.

\bibitem[{\citenamefont{Anselmino and Murgia}(1998)}]{Anselmino:1998yz}
\bibinfo{author}{\bibfnamefont{M.}~\bibnamefont{Anselmino}} \bibnamefont{and}
  \bibinfo{author}{\bibfnamefont{F.}~\bibnamefont{Murgia}},
  \bibinfo{journal}{Phys. Lett.} \textbf{\bibinfo{volume}{B442}},
  \bibinfo{pages}{470} (\bibinfo{year}{1998}), \eprint{hep-ph/9808426}.

\bibitem[{\citenamefont{Anselmino et~al.}(1999)\citenamefont{Anselmino,
  Boglione, and Murgia}}]{Anselmino:1999pw}
\bibinfo{author}{\bibfnamefont{M.}~\bibnamefont{Anselmino}},
  \bibinfo{author}{\bibfnamefont{M.}~\bibnamefont{Boglione}}, \bibnamefont{and}
  \bibinfo{author}{\bibfnamefont{F.}~\bibnamefont{Murgia}},
  \bibinfo{journal}{Phys. Rev.} \textbf{\bibinfo{volume}{D60}},
  \bibinfo{pages}{054027} (\bibinfo{year}{1999}),
  \eprint[http://arXiv.org/abs]{hep-ph/9901442}.

\bibitem[{\citenamefont{Boglione and Mulders}(2000)}]{Boglione:2000jk}
\bibinfo{author}{\bibfnamefont{M.}~\bibnamefont{Boglione}} \bibnamefont{and}
  \bibinfo{author}{\bibfnamefont{P.~J.} \bibnamefont{Mulders}},
  \bibinfo{journal}{Phys. Lett.} \textbf{\bibinfo{volume}{B478}},
  \bibinfo{pages}{114} (\bibinfo{year}{2000}),
  \eprint[http://arXiv.org/abs]{hep-ph/0001196}.

\bibitem[{\citenamefont{Efremov et~al.}(2001)\citenamefont{Efremov, Goeke, and
  Schweitzer}}]{Efremov:2001cz}
\bibinfo{author}{\bibfnamefont{A.~V.} \bibnamefont{Efremov}},
  \bibinfo{author}{\bibfnamefont{K.}~\bibnamefont{Goeke}}, \bibnamefont{and}
  \bibinfo{author}{\bibfnamefont{P.}~\bibnamefont{Schweitzer}},
  \bibinfo{journal}{Phys. Lett.} \textbf{\bibinfo{volume}{B522}},
  \bibinfo{pages}{37} (\bibinfo{year}{2001}), \bibinfo{note}{erratum-ibid.\
  {\bf{B544}} (2002) 389}, \eprint[http://arXiv.org/abs]{hep-ph/0108213}.

\bibitem[{\citenamefont{Bacchetta et~al.}(2001)\citenamefont{Bacchetta, Kundu,
  Metz, and Mulders}}]{Bacchetta:2001di}
\bibinfo{author}{\bibfnamefont{A.}~\bibnamefont{Bacchetta}},
  \bibinfo{author}{\bibfnamefont{R.}~\bibnamefont{Kundu}},
  \bibinfo{author}{\bibfnamefont{A.}~\bibnamefont{Metz}}, \bibnamefont{and}
  \bibinfo{author}{\bibfnamefont{P.~J.} \bibnamefont{Mulders}},
  \bibinfo{journal}{Phys. Lett.} \textbf{\bibinfo{volume}{B506}},
  \bibinfo{pages}{155} (\bibinfo{year}{2001}),
  \eprint[http://arXiv.org/abs]{hep-ph/0102278}.

\bibitem[{\citenamefont{Bacchetta et~al.}(2002)\citenamefont{Bacchetta, Kundu,
  Metz, and Mulders}}]{Bacchetta:2002tk}
\bibinfo{author}{\bibfnamefont{A.}~\bibnamefont{Bacchetta}},
  \bibinfo{author}{\bibfnamefont{R.}~\bibnamefont{Kundu}},
  \bibinfo{author}{\bibfnamefont{A.}~\bibnamefont{Metz}}, \bibnamefont{and}
  \bibinfo{author}{\bibfnamefont{P.~J.} \bibnamefont{Mulders}},
  \bibinfo{journal}{Phys. Rev.} \textbf{\bibinfo{volume}{D65}},
  \bibinfo{pages}{094021} (\bibinfo{year}{2002}), \eprint{hep-ph/0201091}.

\bibitem[{\citenamefont{Brodsky
  et~al.}(2002{\natexlab{a}})\citenamefont{Brodsky, Hwang, and
  Schmidt}}]{Brodsky:2002cx}
\bibinfo{author}{\bibfnamefont{S.~J.} \bibnamefont{Brodsky}},
  \bibinfo{author}{\bibfnamefont{D.~S.} \bibnamefont{Hwang}}, \bibnamefont{and}
  \bibinfo{author}{\bibfnamefont{I.}~\bibnamefont{Schmidt}},
  \bibinfo{journal}{Phys. Lett.} \textbf{\bibinfo{volume}{B530}},
  \bibinfo{pages}{99} (\bibinfo{year}{2002}{\natexlab{a}}),
  \eprint[http://arXiv.org/abs]{hep-ph/0201296}.

\bibitem[{\citenamefont{Collins}(2002)}]{Collins:2002kn}
\bibinfo{author}{\bibfnamefont{J.~C.} \bibnamefont{Collins}},
  \bibinfo{journal}{Phys. Lett.} \textbf{\bibinfo{volume}{B536}},
  \bibinfo{pages}{43} (\bibinfo{year}{2002}), \eprint{hep-ph/0204004}.

\bibitem[{\citenamefont{Ji and Yuan}(2002)}]{Ji:2002aa}
\bibinfo{author}{\bibfnamefont{X.}~\bibnamefont{Ji}} \bibnamefont{and}
  \bibinfo{author}{\bibfnamefont{F.}~\bibnamefont{Yuan}},
  \bibinfo{journal}{Phys. Lett.} \textbf{\bibinfo{volume}{B543}},
  \bibinfo{pages}{66} (\bibinfo{year}{2002}), \eprint{hep-ph/0206057}.

\bibitem[{\citenamefont{Belitsky et~al.}(2003)\citenamefont{Belitsky, Ji, and
  Yuan}}]{Belitsky:2002sm}
\bibinfo{author}{\bibfnamefont{A.~V.} \bibnamefont{Belitsky}},
  \bibinfo{author}{\bibfnamefont{X.}~\bibnamefont{Ji}}, \bibnamefont{and}
  \bibinfo{author}{\bibfnamefont{F.}~\bibnamefont{Yuan}},
  \bibinfo{journal}{Nucl. Phys.} \textbf{\bibinfo{volume}{B656}},
  \bibinfo{pages}{165} (\bibinfo{year}{2003}), \eprint{hep-ph/0208038}.

\bibitem[{\citenamefont{Boer et~al.}(2003{\natexlab{a}})\citenamefont{Boer,
  Mulders, and Pijlman}}]{Boer:2003cm}
\bibinfo{author}{\bibfnamefont{D.}~\bibnamefont{Boer}},
  \bibinfo{author}{\bibfnamefont{P.~J.} \bibnamefont{Mulders}},
  \bibnamefont{and} \bibinfo{author}{\bibfnamefont{F.}~\bibnamefont{Pijlman}}
  (\bibinfo{year}{2003}{\natexlab{a}}), \eprint{hep-ph/0303034}.

\bibitem[{\citenamefont{Brodsky
  et~al.}(2002{\natexlab{b}})\citenamefont{Brodsky, Hwang, and
  Schmidt}}]{Brodsky:2002rv}
\bibinfo{author}{\bibfnamefont{S.~J.} \bibnamefont{Brodsky}},
  \bibinfo{author}{\bibfnamefont{D.~S.} \bibnamefont{Hwang}}, \bibnamefont{and}
  \bibinfo{author}{\bibfnamefont{I.}~\bibnamefont{Schmidt}},
  \bibinfo{journal}{Nucl. Phys.} \textbf{\bibinfo{volume}{B642}},
  \bibinfo{pages}{344} (\bibinfo{year}{2002}{\natexlab{b}}),
  \eprint{hep-ph/0206259}.

\bibitem[{\citenamefont{Boer et~al.}(2003{\natexlab{b}})\citenamefont{Boer,
  Brodsky, and Hwang}}]{Boer:2002ju}
\bibinfo{author}{\bibfnamefont{D.}~\bibnamefont{Boer}},
  \bibinfo{author}{\bibfnamefont{S.~J.} \bibnamefont{Brodsky}},
  \bibnamefont{and} \bibinfo{author}{\bibfnamefont{D.~S.} \bibnamefont{Hwang}},
  \bibinfo{journal}{Phys. Rev.} \textbf{\bibinfo{volume}{D67}},
  \bibinfo{pages}{054003} (\bibinfo{year}{2003}{\natexlab{b}}),
  \eprint{hep-ph/0211110}.

\bibitem[{\citenamefont{Gamberg
  et~al.}(2003{\natexlab{a}})\citenamefont{Gamberg, Goldstein, and
  Oganessyan}}]{Gamberg:2003ey}
\bibinfo{author}{\bibfnamefont{L.~P.} \bibnamefont{Gamberg}},
  \bibinfo{author}{\bibfnamefont{G.~R.} \bibnamefont{Goldstein}},
  \bibnamefont{and} \bibinfo{author}{\bibfnamefont{K.~A.}
  \bibnamefont{Oganessyan}}, \bibinfo{journal}{Phys. Rev.}
  \textbf{\bibinfo{volume}{D67}}, \bibinfo{pages}{071504}
  (\bibinfo{year}{2003}{\natexlab{a}}), \eprint{hep-ph/0301018}.

\bibitem[{\citenamefont{Metz}(2002)}]{Metz:2002iz}
\bibinfo{author}{\bibfnamefont{A.}~\bibnamefont{Metz}}, \bibinfo{journal}{Phys.
  Lett.} \textbf{\bibinfo{volume}{B549}}, \bibinfo{pages}{139}
  (\bibinfo{year}{2002}), \eprint[http://arXiv.org/abs]{hep-ph/0209054}.

\bibitem[{\citenamefont{Gamberg
  et~al.}(2003{\natexlab{b}})\citenamefont{Gamberg, Goldstein, and
  Oganessyan}}]{Gamberg:2003eg}
\bibinfo{author}{\bibfnamefont{L.~P.} \bibnamefont{Gamberg}},
  \bibinfo{author}{\bibfnamefont{G.~R.} \bibnamefont{Goldstein}},
  \bibnamefont{and} \bibinfo{author}{\bibfnamefont{K.~A.}
  \bibnamefont{Oganessyan}} (\bibinfo{year}{2003}{\natexlab{b}}),
  \eprint{hep-ph/0307139}.

\bibitem[{\citenamefont{Levelt and Mulders}(1994)}]{Levelt:1994np}
\bibinfo{author}{\bibfnamefont{J.}~\bibnamefont{Levelt}} \bibnamefont{and}
  \bibinfo{author}{\bibfnamefont{P.~J.} \bibnamefont{Mulders}},
  \bibinfo{journal}{Phys. Lett.} \textbf{\bibinfo{volume}{B338}},
  \bibinfo{pages}{357} (\bibinfo{year}{1994}),
  \eprint[http://arXiv.org/abs]{hep-ph/9408257}.

\bibitem[{\citenamefont{Mulders and Tangerman}(1996)}]{Mulders:1996dh}
\bibinfo{author}{\bibfnamefont{P.~J.} \bibnamefont{Mulders}} \bibnamefont{and}
  \bibinfo{author}{\bibfnamefont{R.~D.} \bibnamefont{Tangerman}},
  \bibinfo{journal}{Nucl. Phys.} \textbf{\bibinfo{volume}{B461}},
  \bibinfo{pages}{197} (\bibinfo{year}{1996}), \bibinfo{note}{erratum-ibid.\
  {\bf B484} (1996) 538}, \eprint[http://arXiv.org/abs]{hep-ph/9510301}.

\bibitem[{\citenamefont{Collins and Soper}(1982)}]{Collins:1982uw}
\bibinfo{author}{\bibfnamefont{J.~C.} \bibnamefont{Collins}} \bibnamefont{and}
  \bibinfo{author}{\bibfnamefont{D.~E.} \bibnamefont{Soper}},
  \bibinfo{journal}{Nucl. Phys.} \textbf{\bibinfo{volume}{B194}},
  \bibinfo{pages}{445} (\bibinfo{year}{1982}).

\bibitem[{\citenamefont{Collins}(2003)}]{Collins:2003fm}
\bibinfo{author}{\bibfnamefont{J.~C.} \bibnamefont{Collins}},
  \bibinfo{journal}{Acta Phys. Polon.} \textbf{\bibinfo{volume}{B34}},
  \bibinfo{pages}{3103} (\bibinfo{year}{2003}), \eprint{hep-ph/0304122}.

\bibitem[{\citenamefont{Manohar and Georgi}(1984)}]{Manohar:1984md}
\bibinfo{author}{\bibfnamefont{A.}~\bibnamefont{Manohar}} \bibnamefont{and}
  \bibinfo{author}{\bibfnamefont{H.}~\bibnamefont{Georgi}},
  \bibinfo{journal}{Nucl. Phys.} \textbf{\bibinfo{volume}{B234}},
  \bibinfo{pages}{189} (\bibinfo{year}{1984}).

\bibitem[{\citenamefont{Kotzinian and Mulders}(1997)}]{Kotzinian:1997wt}
\bibinfo{author}{\bibfnamefont{A.~M.} \bibnamefont{Kotzinian}}
  \bibnamefont{and} \bibinfo{author}{\bibfnamefont{P.~J.}
  \bibnamefont{Mulders}}, \bibinfo{journal}{Phys. Lett.}
  \textbf{\bibinfo{volume}{B406}}, \bibinfo{pages}{373} (\bibinfo{year}{1997}),
  \eprint[http://arXiv.org/abs]{hep-ph/9701330}.

\bibitem[{\citenamefont{Boer and Mulders}(1998)}]{Boer:1998nt}
\bibinfo{author}{\bibfnamefont{D.}~\bibnamefont{Boer}} \bibnamefont{and}
  \bibinfo{author}{\bibfnamefont{P.~J.} \bibnamefont{Mulders}},
  \bibinfo{journal}{Phys. Rev.} \textbf{\bibinfo{volume}{D57}},
  \bibinfo{pages}{5780} (\bibinfo{year}{1998}),
  \eprint[http://arXiv.org/abs]{hep-ph/9711485}.

\end{thebibliography}

\end{document}